\documentclass[10pt,letterpaper]{article}
\usepackage{graphicx}
\usepackage[final=true]{cogsci}

\usepackage{xcolor}

\usepackage[
  style=apa,
  natbib=true,
  annotation=false,
]{biblatex}
\usepackage{float} 

\title{Human-AI Synergy Supports Collective Creative Search}

\usepackage{authblk}

\author[1]{Chenyi Li}
\author[2]{Raja Marjieh}
\author[1]{Haoyu Hu}
\author[3]{Mark Steyvers}
\author[4,5]{Katherine Collins}
\author[6]{Ilia Sucholutsky$^{*}$}
\author[1]{Nori Jacoby$^{*}$}

\affil[1]{Cornell University}
\affil[2]{Princeton University}
\affil[3]{University of California, Irvine}
\affil[4]{Massachusetts Institute of Technology}
\affil[5]{University of Cambridge}
\affil[6]{New York University}

\begin{document}

\maketitle
\begin{abstract} 
The creation of new ideas and content increasingly relies on collectives: not just of multiple humans, but humans and AI agents together.  We study collective generation of new ideas using a controlled word-guessing task that balances open-endedness with an objective measure of task performance. Participants attempt to infer a hidden target word, scored based on the semantic similarity of their guesses to the target, while also observing the best guess from previous players. We found that hybrid human–AI groups showed higher performance than both human-only and AI-only groups. Within hybrid groups, both humans and AI agents systematically adjust their strategies relative to single-agent conditions, suggesting higher-order interaction effects, whereby agents adapt to each other's presence. Although some performance benefits can be reproduced through collaboration between heterogeneous AI systems, human–AI collaboration remains superior, underscoring complementary roles in collective creativity. Together, these findings demonstrate the advantages of human–AI synergy in collective intelligence tasks.


\end{abstract}

\section{Introduction}

Rapid advances in generative AI are giving rise to a hybrid society in which the production of new ideas and content increasingly integrates human and artificial contributions, across domains ranging from creative writing \citep{breithaupt2024humans,porter2024ai} to scientific discovery \citep{jumper2021highly,hao2026artificial,rmus2025generating,schmidgall2025agent,gottweis2025towards}. As people engaged in the discovery of innovative ideas rely more heavily on AI tools as thought partners \citep{collins2024building}, understanding AI’s impact on collective outcomes becomes increasingly important \citep{sucholutsky2025using}. While AI assistance can improve individual task performance, it also risks driving excessive homogenization and reducing collective diversity \citep{doshi2024generative}.
Studying human–AI interactions at the collective level remains challenging using traditional methods. This is, in part, because many real-world idea discovery proceses are inherently open-ended and hard to evaluate in controlled settings with well-defined objectives. Moreover, both humans and AI systems adapt their behavior in response to the outputs of others, making it difficult to infer emergent collective dynamics from studies focused solely on isolated individual human–AI interactions \citep{tsvetkova2024new}.
\begin{figure*}[!t]
  \begin{center}
    \centering
    \includegraphics[width=0.74\textwidth]{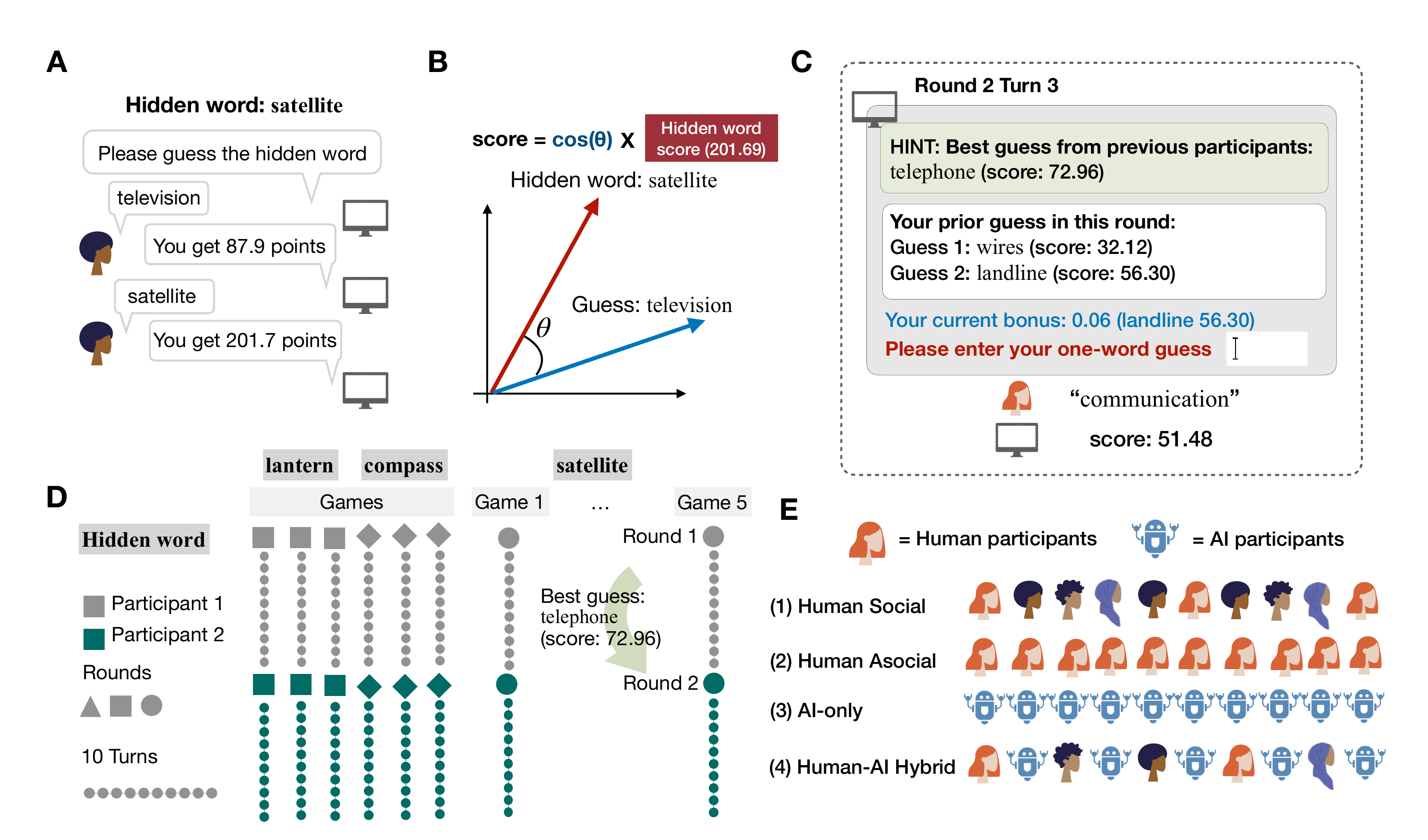}
  \end{center}
  \caption{Experiment framework for collective creative search. (A) Participants attempt to infer a hidden target word (``satellite'') and receive similarity score feedback over multiple turns. (B) The similarity score of each guessed word is computed by the product of the hidden word score and the cosine similarity between them. (C) In each round, the participants received the best guess from previous rounds as a hint. (D) Participants were embedded in a collective guessing game with a chain-like network, where best-guess information was transmitted within each game (chain). Each game had 10 rounds of 10 turns each, totaling 100 guesses per game. (E) Schematic of the different experimental conditions considered.}
  \label{figure-schematics}
\end{figure*}

Here, we address these challenges by studying a creative discovery task with an objective ground truth. Inspired by the game Semantle \citep{semantle2022,ueshima2024simple}, participants infer a hidden target word (Fig. \ref{figure-schematics}A) and receive feedback based on the semantic similarity between their guesses and the target (Fig. \ref{figure-schematics}B). In each round, participants are informed of the best guess and score produced by previous players within the same game (Fig. \ref{figure-schematics}C). Players, either human or Gemini 2.5 AI agents, join groups composed of either humans, AI, or both (Fig. \ref{figure-schematics}D,E). We found that hybrid human-AI groups consistently outperformed homogeneous groups, achieving faster convergence to high-value solutions. Crucially, participation in hybrid groups altered the behaviors of both humans and AI. AI agents exhibited greater lexical diversity and search quality when interacting with humans, while humans achieved slightly higher performance and made significantly more unique guesses in the presence of AI. This mutual adaptation suggests that the benefits of human-AI collaboration emerge not from simple addition but from the dynamic interplay of complementary cognitive strategies: humans explore broadly, preventing premature convergence, while AI exploits efficiently, accelerating progress toward promising regions. To test whether these benefits arise from cognitive heterogeneity rather than agent diversity per se, we analyzed an experiment involving two distinct LLMs, Gemini 2.5 and GPT-5.1. This experiment showed a similar synergistic effect, but overall performance was lower than in human-AI collaboration, indicating that part of the hybrid advantage originates from agent heterogeneity. Finally, we validated in a series of control studies that these effects are robust to different types of AI systems, different forms of communication channels between participants in the game and variations in model decoding temperatures.

\section{Background}
The ability of humans to collectively discover new ideas and create novel content is one of the definitive features of human culture \citep{henrich2015secret,boyd1996culture,tomasello2009cultural}. Research on social learning has shown that innovation depends on factors such as network topology, interaction structure, and group size \citep{mason2008propagation,derex2019causal,shirado2020network,brackbill2020impact}.
However recently AI is increasingly used to assist people in complex cognitive interactions \citep{collins2024building}.  What is the impact of AI assistance on idea discovery across groups of people and AI agents?
\citet{doshi2024generative} examined humans assisted by AI agents in a creative writing task and found that AI assistance enhanced individual creative performance but reduced collective diversity, leading to excessive homogenization. However, their study examined human–AI interaction in isolation and did not directly test the emergent collective phenomena that arise from interactions among multiple humans and AI agents \citep{tsvetkova2024new,brinkmann2023machine}.

One promising medium for studying collective behavior comes from the information foraging literature \citep{garg2022individual}. The idea is that optimal search in any environment requires balancing exploration of new possibilities with exploitation of known resources \citep{hills2015exploration}. This exploration-exploitation tradeoff is a fundamental principle observed across domains, from animal foraging to human memory search \citep{hills2012optimal}. In this literature, it is often noted that individual strategies of exploration and exploitation are similar to the same dynamics at the collective level \citep{hills2015exploration}.

In a recent paper, \citet{ueshima2024simple} used a word guessing game to probe collective creativity in a semantic search setting. They found that simple bots aggregating human guesses had a modest positive effect on group performance for easier words.
A similar beneficial effect of simple bots was reported by \citet{shirado2017locally}, who showed that, in a coordination game, introducing bot agents that inject noise can improve collective coordination by preventing groups from becoming trapped in local minima.
The paper by \citet{shirado2017locally} and the semantic search task by \citet{ueshima2024simple} provide a valuable framework but do not study realistic AI systems. This limitation is important because \citet{tsvetkova2024new} have shown that human–AI interaction can give rise to substantial second-order effects, whereby humans systematically alter their behavior in response to AI agents. These dynamics imply that studying collective behavior requires experiments involving real humans interacting with real AI systems under shared conditions \citep{sucholutsky2025using, hu2026human}.
Recent technological advances in large-scale online experimentation \citep{almaatouq2021empirica,harrison2020gsp,marjieh2025characterizing} now make it possible to embed both humans and real AI agents within large social networks where they interact directly with one another. For example, a recent study embedded humans and AI in hybrid human–AI social networks engaged in a creative writing task \citep{shiiku2025dynamics}. While AI agents initially outperformed humans on the creative task, their output diversity deteriorated over time. In contrast, hybrid human–AI groups began with lower performance but gradually improved, achieving a balance between performance and diversity.
Together, these findings suggest that it is now feasible to study human–AI interaction at scale, under controlled conditions, and with realistic contemporary AI systems, while capturing emergent collective dynamics that cannot be inferred from individual-level studies alone.

\section{Method}

We conducted an online experiment on collective creative discovery, wherein participants (either human or AI) played a word-guessing game with one hidden target word per game (Fig. \ref{figure-schematics}A). Ten target words were selected to span a range of frequencies (and difficulty levels) : `harbor', `door', `pencil', `lantern', `river', `compass', `satellite', `metamorphosis', `topography', `vessel'. Each game consisted of 10 rounds, with 10 guesses per round, yielding 100 guesses per game. Each batch contained 5 games per target word, for 50 games per batch (Fig. \ref{figure-schematics}D); we ran 4 batches, yeilding 200 games in total. All the measures were averaged over 4 batches.

In each round, participants received the best guess from the previous round in the game (Fig. \ref{figure-schematics}C). Participants took ten turns of guesses and received a similarity score after each guess. The guess with the highest score was recorded as a bonus and passed to the next round. 
The similarity score was computed as the product of the cosine similarity between Word2Vec \citep{mikolov2013efficient} embeddings of the guessed and hidden words, scaled by an arbitrary constant (201.69; Fig. \ref{figure-schematics}B). We ensured that the maximum possible score was not revealed to participants, so that they would have an incentive to continue exploring even after identifying the target word. From the original vocabulary of approximately three million tokens, we retained 663,273 common English words after lowercasing and excluding acronyms and brand names that lacked lowercase representations in the model. Guesses that did not appear in this vocabulary were assigned a score of zero.

\subsection{Experimental conditions}
We compared four main experimental conditions (Fig. \ref{figure-schematics}E): (1) \textbf{Human Social ($N=210$)}: all rounds were occupied by human participants, with a different participant playing each round; participants could take part in up to 10 games (one per target word), playing only one round per game. (2) \textbf{Human Asocial ($N=179$)}: each participant was assigned to a single game and completed all 10 rounds, totaling 100 guesses.
(3) \textbf{AI-only} ($20,000$ calls to the Google API): 
all $2,000$ rounds were simulated using Gemini 2.5 Flash. AI agents independently received the same prompt and feedback as humans participants, including the hint, within-round guessing history, and prompt ``Please enter your one-word guess''. 
(4) \textbf{Human-AI Hybrid} ($N=114$, $10,160$ Google API calls): within each game, each round was randomly assigned to either a human participant or a Gemini 2.5 Flash agent, resulting in an average of 5.1 human and 4.9 AI rounds per game ($SD = 1.54$ and $1.55$, respectively; 1058 human and 1018 AI rounds in total). As in the Human Social condition, human participants completed only one round within any given game and participated in 10 games. To minimize bias, participants were not informed that any guesses may have been generated by AI agents.

\subsection{Participants and AI queries}

\textbf{Human participants.} $503$ participants were recruited successfully from Prolific ($237, 242,$ and $24$ self-reported as female, male, and other, respectively; mean age = 38.13, SD = 12.30). Participants were based in the US and identified English as their native language. Participants provided consent under a Cornell University approved protocol (IRB0148995) and were compensated at a rate of \$9 per hour. We conducted a subsampling-based power analysis using the performance difference between the Social and Asocial conditions, a metric independent of the AI and Hybrid conditions, and found that 60 participants per group were required for this effect to reach significance at the $p < .05$ threshold for over  $80 \%$ among  the $1000$ iterations. Our actual sample sizes ($N_{social} = 210$, $N_{asocial} = 179$) comfortably exceed this requirement.

 \textbf{AI queries.} We conducted 165,140 API queries to Gemini 2.5 Flash (version released June 17, 2025) with temperature 0.7. We also used 10,020 API queries to GPT 5.1 for control experiments (version released November 13, 2025) with temperature 0.7, medium reasoning effort. For strategy annotation, we used 8,461 API queries to Claude Sonnet 4 (version released May 22, 2025) with temperature 0.1.
All experiments were conducted using PsyNet \citep{harrison2020gsp}, a Python-based framework for large-scale online psychological experiments and human-AI interactions.

\section{Results}
\begin{figure}[!t]
  \begin{center}
    \centering
    \includegraphics[width=0.45\textwidth]{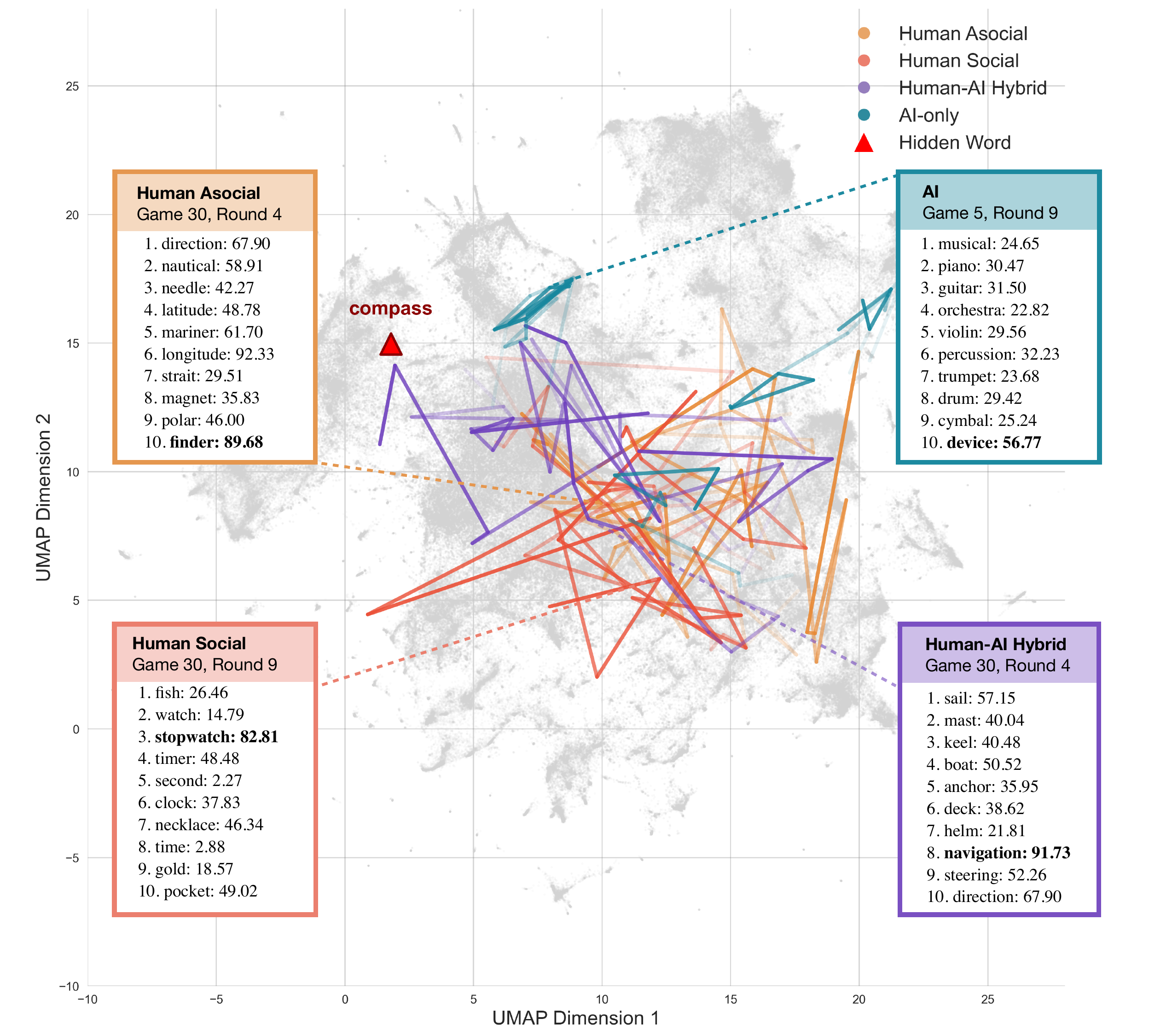}
  \end{center}
  \caption{Semantic exploration trajectories. All words were embedded with a Word2Vec model and projected to a two-dimensional space using UMAP. We present one target word (``compass''). Colored trajectories show five games for each of the four conditions. For each game and each round, we computed the average coordinates of the 10 guesses and connected these round centroids from rounds 1 to 10; line opacity increases with round index. Inset boxes display example rounds showing all 10 guesses with their similarity scores; best guesses of the rounds are highlighted.}
  \label{semantic-figure}
\end{figure}
\begin{figure*}[!t]
  \begin{center}
    \centering
    \includegraphics[width=\textwidth]{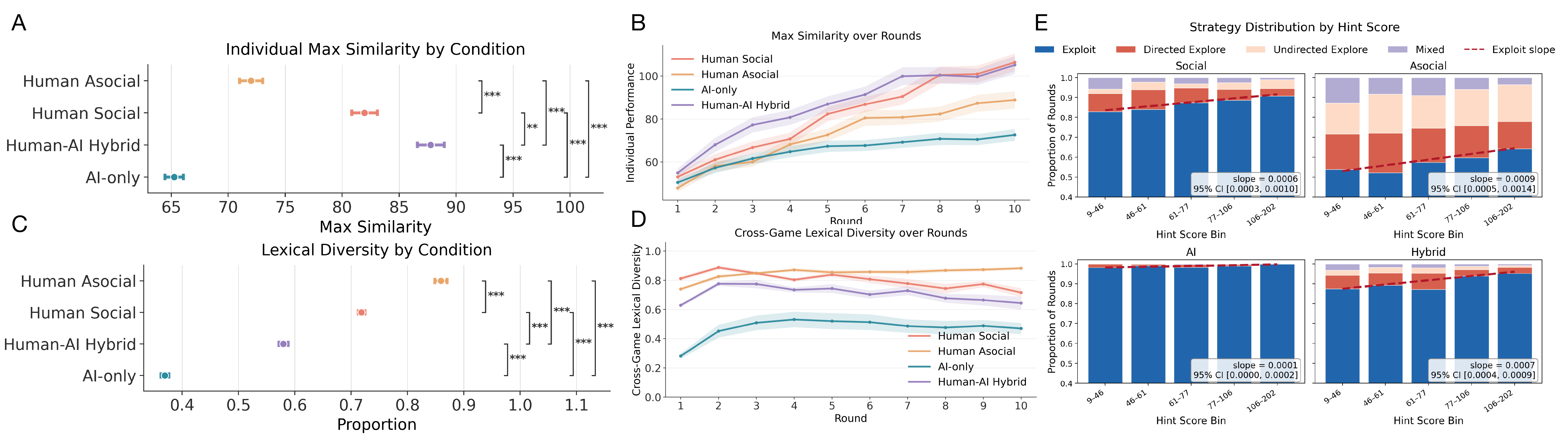}
    
  \end{center}
  \caption{Performance, diversity and strategy. \textbf{A.} Performance, computed as the average of the maximal score across rounds.  Error bars represent one standard error across participants. Asterisks *, **, and  *** denote significance levels of 0.05, 0.01, and 0.001, respectively. To account for multiple comparisons (here and in the rest of the paper), only results that remained significant after Bonferroni correction are shown. \textbf{B.} Performance across rounds. Error bars represent standard error across participants. \textbf{C.} Lexical diversity, computed as the average proportion of unique guessed words of each game. \textbf{D.} Cross-game lexical diversity acorss rounds. 
  \textbf{E.} Strategy distribution by hint score quantile across conditions. The dashed red line shows the OLS trend for P(Exploit) across hint score bins; slope and bootstrapped $95\%$ CI are annotated within each panel.
  }
  \label{main-results}
\end{figure*}
\subsection{Exploration trajectories in semantic space}

We begin by examining exploration trajectories across conditions. We constructed a semantic space using the same Word2Vec embeddings that were used for scoring (see Methods). 
We then applied UMAP \citep{mcinnes2018umap} to reduce these embeddings to two dimensions and plotted them as gray background points. Fig. \ref{semantic-figure} illustrates exploration trajectories for the Human Social, Human Asocial, AI-only, and Human-AI Hybrid conditions in a representative game with the hidden word ``compass''. 
Fig. 2 highlights three key insights. First, trajectories in the AI-only condition form compact clusters in semantic space; for ``compass'', these clusters center around word groups such as music instrument. Second, the Human Social and Human Asocial conditions show more widely spread trajectories, with higher semantic diversity than the AI condition. Third, the Human-AI Hybrid condition exhibits similarly expansive trajectories and is generally closer to the hidden word ``compass'' than the other conditions.

\subsection{Performance analysis}

To measure \textbf{performance} in creative search, we took the maximum similarity score reached by each round in the game and averaged these maxima across all rounds in all games in each condition. 
As shown in Fig.~\ref{main-results}A, the Human-AI Hybrid condition produced the highest performance  while the AI-only condition produced the lowest. The Hybrid condition yielded significantly higher similarity scores compared with the other conditions ($p<.001$). Human Social surpassed Human Asocial conditions significantly ($p<.001$), showing the advantage of collective behaviors which is our main focus. 

We next analyzed  performance as a function of round within the game (Fig.~\ref{main-results}B). Across all conditions, scores increased significantly over time ($p<.001$); however, the magnitude of improvement differed by condition, with AI showing smaller gains ($\Delta = 22.10$) than the other groups ($\Delta = 40.93-53.21$).
The dynamics also revealed that different conditions reached  peak performance at different stages. The Human-AI Hybrid condition exhibited a consistently increasing trajectory throughout the game, maintaining the highest scores in the early and middle rounds. Notably, it surged rapidly in the early stage: by round 3, it already showed significantly higher maximum similarity than both the Human Asocial and AI-only conditions ($p<.005$). The AI-only condition showed modest early gains but plateaued after round 5, ultimately achieving the lowest final scores. In contrast, the Human Social condition showed a later surge, with performance accelerating sharply around rounds 7–8 and converging with the Hybrid condition by rounds 9–10.

\subsection{Diversity analysis}
To measure \textbf{diversity}, we used \textbf{lexical diversity}, defined as the average proportion of unique guessed words within each game (Fig.~\ref{main-results}C). The Human Asocial condition exhibited the highest diversity ($M=0.86$), likely reflecting higher exploration rates or greater noise. In contrast, the AI-only condition showed markedly lower diversity ($M=0.37$), indicating a tendency to remain confined to a narrow region of semantic space despite consistently receiving relatively low scores. The Human-AI Hybrid condition showed reduced lexical diversity ($M=0.58$) compared with the human conditions, but significantly higher diversity than the AI-only condition ($p<.001$). Interestingly, Fig.~\ref{main-results}D shows that both the Human Social and Human-AI Hybrid conditions decreased in diversity across iterations, likely because guesses became closer to the target over time. These results suggest that Hybrid Human-AI groups occupy an intermediate position in terms of diversity, preserving more exploratory variation than AI-only groups while reducing the high variability observed in human-only settings.

\subsection{Guessing strategy analysis}
To characterize guessing strategies beyond vocabulary diversity, we implemented an LLM-as-a-judge annotation pipeline using Claude Sonnet 4. In a preliminary step, an independent instance of the model reviewed the full set of semantic guess trajectories across the main conditions and inductively derived seven strategy types defined relative to the hint. The annotation model then classified each round of 10 guesses into a soft probability distribution over the seven types. Each round was then assigned a discrete label by taking the strategy with the highest probability, and the seven labels were further collapsed into four broader categories: Exploit (same-category exploit or subcategory shift), Directed Explore (associative or inferential category break), Undirected Explore (random category break or hint-independent), and Mixed (mixed strategy). 

Fig.~\ref{main-results}D shows how the strategy distribution varied over five equal-frequency hint score quantiles across conditions. A red dashed ordinary least squares (OLS) trend line for P(Exploit) is overlaid, with the bootstrapped slope and 95\% CI annotated. Both human conditions demonstrated clear adaptive behaviors (Human Social: $M = 0.0006$, 
$\text{CI}_{95\%}=$ $[0.0003,0.0010]$; Human Asocial: $M = 0.0011$, 
$\text{CI}_{95\%}=$ $[0.0005,0.0014]$): when hint scores were low, they relied more on Explore or Mixed strategies to escape unpromising regions of semantic space; as hint quality increased, they progressively shifted towards Exploit strategies. In contrast, the AI-only condition showed virtually no adaptation to hint quality ($M = 0.0001$, 
$\text{CI}_{95\%}=$ $[0.0000,0.0002]$). Exploit was the dominant strategies ($99\%$) in AI-only group, higher than that of other conditions ($p<.001$). This indicates that AI tends to focus on nearby semantic region of the hint and lacks the adaptivity to escape local maxima. The Human-AI Hybrid condition displayed an intermediate pattern. Its strategy distribution was more adaptive than the AI-only condition ($p=.0036$) and achieved compatative adaptivity to the hint score, compared to Human Social ($p=1.0$) and Human Asocial ($p=1.0$). 

\subsection{Interaction with people changes AI behavior}

\begin{figure}[!t]
  \begin{center}
    \centering
    \includegraphics[width=0.36\textwidth]{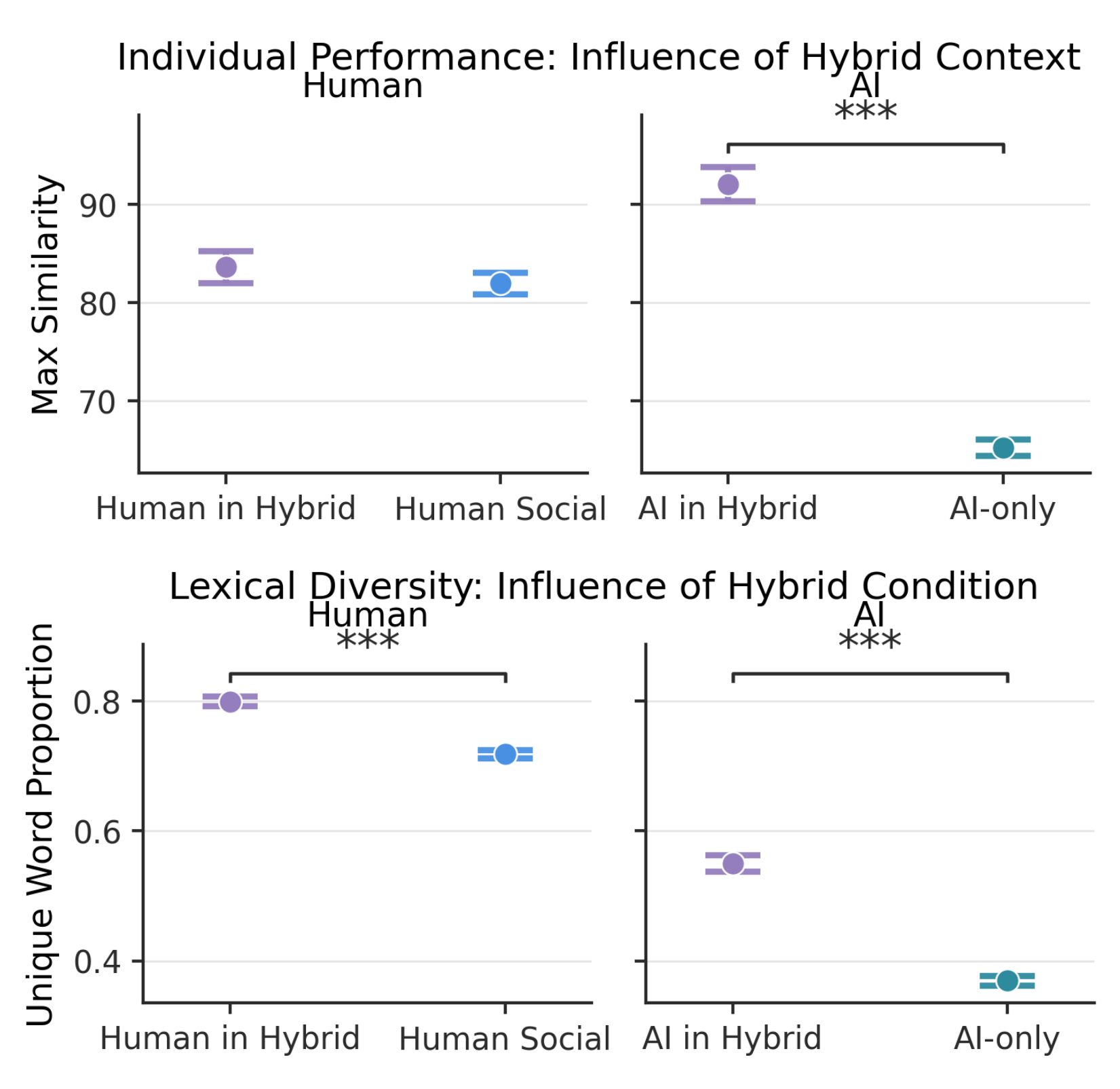}
  \end{center}
  \caption{Indirect influence of AI on human behavior and of humans on AI in the Human-AI Hybrid condition. \textbf{A.} Performance: comparison of human performance in the purely human and Hybrid (Human–AI) conditions, and AI performance in the purely AI and Hybrid conditions. \textbf{B.} Same for lexical diversity.}
  \label{figure-indirect-hybrid}
\end{figure}

Previous theoretical work \citep{tsvetkova2024new} suggests that humans and AI can influence each other indirectly by observing and responding to each other’s behavior (second-order effects).
To quantify the influence of Hybrid Human-AI condition on both human and AI agents, we compared their performance and diversity. Fig. \ref{figure-indirect-hybrid}A shows that human participants in the Hybrid condition did not differ significantly in individual performance from the Human Social group ($p=0.392, d=0.032$). In contrast, AI agents in the Hybrid condition performed significantly better than in the AI-only condition ($p<.001, d=0.618$). These results suggest that AI agents, in particular, modified their behavior when exposed to input from the other type of agent. 

As shown in Fig. \ref{figure-indirect-hybrid}B, humans in the Hybrid condition contributed a significantly higher proportion of unique words (\textbf{lexical diversity}) compared to the Human Social condition ($p < .001$). Similarly, AI agents in the Hybrid condition also showed a significantly higher lexical diversity compared to the AI-only condition ($p < .001$).

Taken together, these results reveal an asymmetric complementarity between human and AI agents. AI agents benefited most from hybrid condition: human input helped them shift toward more varied and productive regions of semantic space. In contrast, human showed little change in performance or diversity, suggesting they served primarily as a source of exploratory signal that guided AI toward more promising regions of semantic space.

\subsection{Controlling for effect of model type and synergy}

To understand the effect of different agents, we implemented a Hybrid AI condition consisting of Gemini 2.5 Flash (50\%) and GPT 5.1 (50\%) agents, and compared it to our original experiment with Gemini 2.5 Flash and new experiment with GPT 5.1 alone. This Hybrid AI condition ($M = 72.83$, $\text{CI}_{95\%}=$ $[70.87,74.80]$) showed significantly higher maximal similarity score than Gemini 2.5 (AI-only) condition ($M = 65.25$, $\text{CI}_{95\%}=$ $[63.66, 66.84]$; $t(3998) = 5.88$, $p < .001$, $d = 0.19$) and the GPT 5.1 condition ($M = 64.08$, 
$\text{CI}_{95\%}=$ $[62.59, 65.58]$; $t(3998) = 6.94$, $p < .001$, $d = 0.22$) but remained significantly worse than the Human-AI Hybrid condition ($M = 87.76$; $t(4081) = -14.93$, $p < .001$, 
$d = 0.30$). 

For diversity, Gemini 2.5 (AI-only condition) displayed similar lexical diversity compared to Hybrid AI ($M = 0.37$, $\text{CI}_{95\%}=$ $[0.35,0.38]$, $p =1.0$, $d=0.017$) and higher diversity than GPT 5.1 ($M = 0.29$, $\text{CI}_{95\%}=$ $[0.28, 0.31]$, $p < .001$, $d=0.76$). 


\subsection{Controlling for prompts and social information}

We conducted additional AI-only control experiments to assess the robustness of our main findings to variation in prompt design, social information format, and model decoding temperature. 

Three alternative prompt design and social information conditions yielded no significant performance differences relative to the main condition (all $p>.1$,$|d| \leq 0.098$): a Complete History condition, in which agents received the full guessing history of previous participants; a Short Advice condition, in which agents received and passed on a single-word piece of advice; and a Long Advice condition, in which agents exchanged multi-sentence natural-language advice. Together, these results suggest that the best-guess signal used in the main experiment provides a sufficient and efficient mechanism for transmitting progress across rounds.




We further varied the decoding temperature ($0.3,0.5,0.9,1.1$); none of these variants outperformed the main condition (all $|d|\leq 0.12$), and although higher temperature modestly increased lexical diversity, it did not translate into performance gains. The results suggest that increasing output randomness alone is insufficient to overcome AI's tendency toward local exploitation. 

Overall, the control experiments validate that our main results are robust to variation in prompt design, transmitted social information and moel decoding temperature.

\section{Discussion}
This study examined how advanced large language models (LLMs) shape collective creative search when embedded within human–AI collaborative groups. We found that hybrid human–AI groups consistently outperformed both human-only and AI-only groups while obtaining intermediate diversity and maintaining  an adaptive strategy. In contrast, AI-only groups exhibited lower overall performance, and this disadvantage proved robust across different AI systems, prompting strategies, and forms of social information (e.g., single-word versus sentence-level feedback).

Our results highlight how humans, AI agents, and hybrid collectives employ fundamentally different search strategies. Human groups prioritize exploration and maintain diversity, but face limitations in rapidly exploiting promising semantic regions. By contrast, AI-only groups focused on narrow regions of semantic space and achieved relatively high average similarity scores early on; however, despite this efficiency, they frequently failed to make decisive breakthroughs to the hidden targets and became trapped in local optima, likely due to their reliance on exploit strategies regardless of hint quality.
These findings align with prior work on complementary inductive biases in humans and AI systems \citep{steyvers2022bayesian}: humans excel at broad, expansive exploration, whereas AI systems, at present, excel at focused exploitation. Hybrid human–AI group combines these strengths. Human exploration introduces semantically diverse guesses that disrupt AI's tendency toward narrow convergence, preventing premature fixation on suboptimal semantic regions. In turn, the concentrated and high-quality guesses generated by AI sharpen human exploration and accelerate movement toward highly relevant areas of semantic space. Moreover, we found that in hybrid settings, humans exhibited increased exploratory diversity, while AI agents displayed greater performance and diversity than in AI-only group. This mutual adaptation suggests that creative performance in hybrid systems emerges through co-adaptation between complementary cognitive strengths. Aligned with this, some—but not all—of these benefits can be partially replicated by combining heterogeneous AI systems from different providers (e.g., Google and OpenAI).

However, there are several limitations to our current work. First, our task operationalizes the complex process of creating new ideas as semantic search using single-word guesses with computational similarity feedback. While this design enables experimental control, it does not capture the full richness of real-world creative processes such as writing, design, or scientific discovery, which unfold in high-dimensional spaces with open-ended and often multiple objectives. Future work should aim to develop controlled yet richer task environments as testbeds for creativity \citep{shiiku2025dynamics}. Second, we focused on a linear chain  structure and a restricted communication channel that transmitted only best-guess information. These design choices likely shape exploration–exploitation dynamics; future studies should examine richer network topologies (e.g., small-world or fully connected networks; \cite{marjieh2025characterizing}) and alternative information-sharing schemes, including synchronous collaboration. Third, our conclusions regarding AI behavior are conditioned on the specific LLMs and prompting strategies used here. We have only begun to explore different models and training regimes that may yield different convergence and diversity dynamics. Finally, our diversity measure relies on a simple count of word uniqueness that may not fully capture novelty or usefulness; combining these metrics with human evaluations or downstream task performance would strengthen external validity.

In conclusion, this work represents a first step towards an experimental understanding of collective human–AI collaboration. Our findings demonstrate that human and AI benefit from each other, but also highlight the importance of understanding human contributions in terms of emergent collective behavior rather than isolated individual performance. Studying creativity at the collective level is therefore essential for designing and understanding future human and AI hybrid societies~\citep{brinkmann2023machine, collins2025revisiting}.

\section{Acknowledgments}
This work was supported by the NSF grant ``Collaborative Research: Research Infrastructure: HNDS-I: Building Infrastructure to Study Human-AI
Hybrid Societies in Experimental Social Networks'' (Award BCS-2523500) and partially supported by the NSF grant ``Collaborative Research: Designing smart environments to augment collective learning \& creativity'' (BCS-2421386). KMC acknowledges support from the NSF SBE SPRF.

 ChatGPT version 5.2 (OpenAI) was used to assist manuscript editing and proofreading. Authors reviewed each of the edit suggestions, and approved the final version.

\printbibliography

\end{document}